\documentclass[referee]{aa} 
\usepackage{graphicx}
\usepackage{txfonts}
\usepackage{natbib}
\usepackage[normalem]{ulem}
\usepackage{cancel}
\bibpunct{(}{)}{;}{a}{}{,} 
\begin{document}

\title{Fundamental plane: dark matter and dissipation contributions}

   \author{A.L.B. Ribeiro
          \inst{1} and C.C. Dantas \inst{2}}
        
          
   \offprints{A.L.B. Ribeiro}

   \institute{Departamento de Ci\^encias Exatas
e Tecnol\'ogicas\\ Universidade Estadual de Santa Cruz -- 45650-000, Ilh\'eus-BA, Brazil\\
              \email{albr@uesc.br}\\
              \and Divis\~ao de Materiais, Instituto de Aeron\'autica e Espa\c co\\ Departamento de Ci\^encia e Tecnologia Aeroespacial \\ P\c ca. Mal. Eduardo Gomes, 50 -- Vila das Ac\'acias -- 12228-904 --S\~ao Jos\'e dos Campos-SP, Brazil\\
\email{ccdantas@iae.cta.br}}

\date{Received December 31, 2009; Accepted April 23, 2010}

 
  \abstract
   { Stellar and galactic systems are objects in dynamical equilibrium that are composed of ordinary baryonic matter hypothetically embedded in extended dominant dark matter halos.}
   {Our aim is to investigate the scaling relations and dissipational features of these objects over a wide range of their properties, taking the dynamical influence of the dark matter component into account.}
   {We study the physical properties of these self-gravitating systems using the two-component virial theorem in conjunction with data that embrace a wide range of astrophysical systems.}
   {We find that the scaling relations defined by the properties of these objects admit a dark-to-luminous density ratio parameter as a natural requirement in this framework. We also probe dissipational effects on the fundamental surface defined by the two-component virial theorem and discuss their relations with respect to the region devoid of objects in the data distribution.}
   {Our results indicate complementary contributions of dissipation and dark matter to the orign of scaling relations in astrophysical systems.}

\titlerunning{Scaling relations and dissipational features on the 2VT}
   \maketitle
%
\section{Introduction \label{Introd}} 

The present cosmological paradigm of structure formation relies on indirect evidence that a dark matter component dominates the dynamics of large structures \citep[e.g.][and references therein]{Pad06}. Gravitational instabilities increase the initially small amplitude fluctuations in the primordial universe, resulting in the evolution towards virialized systems like galaxies and clusters of galaxies \citep{Pee80,Col95,Dod03}. However, this rather simple mechanism is in reality exceedingly more complicated, mainly because of non-linear effects, such as the role of dissipation of the baryonic component inside dark matter halos \citep{Sil81} and the nature of dark matter itself \citep[e.g.][]{Gao07}. In addition, in the hierarchical scenario of structure formation, low-mass objects are formed first, and then larger systems are formed by subsequent merging of smaller subunits, resulting to a greater or lesser degree in a dynamically complicated history of fusions for probably any given selfgravitating object seen today (e.g., \citealt{DeL07,McI08}).

Despite the evident difficulties reaching complete understanding of these mechanisms, it is remarkable that virialized systems do present some kind of global regularity in their scaling relations. For instance, it is found that all virialized stellar systems are reasonably confined to a two-dimensional sheet in their parameter space defined by three independent variables, derivable from quantities such as luminosity, radius, and projected velocity dispersion \citep[e.g.][]{Sch93,Bur97}. The study of elliptical galaxies, for instance, is of paramount importance in this respect, given that their 2-D manifold of physical observables are a very well-defined plane (the so-called `fundamental plane'', c.f. \citealt{Djo87,Dre87}), with a characteristic small scatter and significant deviation (``tilt'') from the simple virial expectation (see \citealt[and references therein]{Tor09}).

The idea of the fundamental plane (FP) has its origin in some restricted relations between global observables (central velocity dispersion, effective radius, and mean effective surface brightness) in stellar systems that emerge after statistical analysis of extensive photometric and spectral data accumulated for a number of objects over the past two decades \citep{Djo87,Dre87,Djo90,Ber03}. The FP is a bivariate family of correlations that are well established for elliptical galaxies in the optical and infrared ranges (e.g. \citealt{Guz93,Pah95}, respectively). Other FP-like relations are observed for spiral and dwarf galaxies \citep[e.g.][]{Jab96,Cod09} and galaxy clusters \citep{Sch93,Fri99,Lan04,Ara09}. Indeed, \citet{Bur97} propose the term ``cosmic metaplane" to describe this ensemble of FPs from globular to  galactic cluster scale.

The existence of such a metaplane reinforces the idea of some dynamical regularity in the properties of a wide range of different astrophysical objects, which certainly deserves deep analysis of its own. But the cosmic metaplane also reveals some other interesting features of the structure formation process, such as a region that seems to be completely devoid of objects, called the ``zone of exclusion'' (ZOE, e.g.  \citealt{Bur97}). Such a feature should be accounted for by the current paradigm of galaxy formation and clustering in the Universe.

\citet{Dan00} (hereafter Paper I) shed some light on these questions by putting forward a model based on the hypothesis that self-gravitating stellar systems have extended dark matter halos and  described their equilibrium state by a modified, two-component virial theorem (2VT). They have found a fundamental surface in the parameter space that embraces the properties of a wide range of stellar and galactic systems. This surface shows remarkable compatibility with the cosmic metaplane discussed by \citet{Bur97} and could partially explain its existence from simple dynamical arguments. 

It turns out that the 2VT hypothesis, however compelling it may seem considering the wide scope of compatiblity resulting from few fundamental assumptions, does not necessarily constrain or contradict the few other alternative explanations to the FP considered in the literature  (see, e.g., \citealt{Pro08,Tor09} and references therein). Indeed, the nature of scaling relations such as the FP is still controversial at this time. In order to advance our understanding of the 2VT hypothesis as the main factor in explaining the scaling relations of virialized systems, we analyze possible paths to disentangling additional effects from the purely dynamical one as imprinted in the 2VT relations themselves. 

A usual and important additional nondynamical effect to explain the FP is dissipation. For instance, \cite{Rob06}  demonstrate that if disk galaxy mergers
are strongly dissipational, tilt is generated in an FP, in agreement with observational determinations. Actually, some authors consider dissipation as a necessary and {\it sufficient} condition for explaining the nature of the FP \citep[e.g.][]{Hop08}. This possibility motivates an analysis of how far the 2VT hypothesis is orthogonal to the dissipational hypothesis, if at all. For that we will define quantitative estimators for dissipation. The most interesting features to analyze under these estimators is the ZOE, a subject that has not been addressed much in the past $\sim$ 10 years since the publication of Paper I (see \citealt{Zar06,Gad09} for recent examples of the ZOE for spheroids and bulges, respectively). However, in that work, the origin of the ZOE and the gap between stellar and galactic systems was not addressed, so this analysis is offered here.

This paper is organized as follows. In the next section, we review the 2VT hypothesis and possible evidence for it. We outline the interplay between dissipation, the 2VT, and the ZOE in Section \ref{Interplay}. Finally, we discuss our results (Section \ref{Discussion}).

\section{Dark matter: scaling relations and 2VT \label{Re2VT}}

In this section we revisit the 2VT formulation and study the evidence for it based on a statistical analysis of the 2VT fittings to a wide range of astrophysical objects.

\subsection{Revisiting the 2VT \label{Revisit}}

The properties of regular nearby galaxies are slowly evolving at present, after a presumably previous and complex gas-rich protogalactic stage. Thus, observables such as surface brightness, effective radius, mean velocity dispersion, and luminosity can be used to specify the parameters of these objects. At the same time, when considering galaxies as stationary self-gravitating systems, their physical properties should at some level reflect an equilibrium state through the virial relation, $r_e \propto \sigma_0^2I_e^{-1}$ (using the one-component virial theorem, 1VT), which implies that only two of a great number of observables are independent. Actually, the FP describes a global relation of virial type, $r_e \propto \sigma_0^AI_e^{B}$ but, differently from  1VT, with exponents $A\sim 1.53$ and $B \sim -0.79$ for elliptical galaxies observed in the near-infrared \citep{Pah98}. This means that there is no direct mapping between  the virial plane and the FP. The reason for the difference of these relations (also called the FP tilt) is a question of current debate involving different hypotheses such as the non-homology of ellipticals properties or a mass-luminosity ratio variable with the total luminosity (see, for instance, \citealt{Cap95,Cap97,Hjo95,Cio96, Kri97,Bek98}). \cite{Tru04} suggests that the FP tilt could be a combination of stellar population plus non-homology effects, while \citet{Lan05} points out scaling relations of ellipticals might be the combined result of cosmological collapse plus dissipative merging. 

When extending the problem to other scales and systems, we find a similar situation: the cosmic virial plane described by the 1VT, defined as the ensemble of all possible collapsed objects of all masses and radii \citep{Bur97}, does not coincide with the observationally determined cosmic metaplane, as shown in Paper I. This metaplane is also tilted with respect to the virial expectations. This problem is mitigated in Paper I by adding a term to the virial equilibrium equation considering the gravitational energy due to the interaction of  dark and baryonic components. Applying the 2VT to stellar and galaxy systems, they find 
\begin{equation}
\langle v^2 \rangle = {GM_{LUM}\over r_{LUM}} + {4\pi\over 3}G\rho_{DM}\langle r^2 \rangle_{LUM}, \label{Eq2VT}
\end{equation}
\noindent where $\rho_{DM}$  is the mean density of the dark matter halo within 
the region containing the luminous component, $\langle v^2 \rangle_{LUM}$ is the
mean square velocity of the stars (or galaxies) and
\begin{equation}
\langle r^2 \rangle_{LUM}\equiv {\int r^2\rho_{LUM}(r)dV\over \int\rho_{LUM}(r)dV},
\end{equation}
\noindent where $\rho_{LUM}$ is the baryonic density inside $r_{LUM}$. In terms of observed quantities, Eq. \ref{Eq2VT} amounts to
\begin{equation}
\sigma_0^2 = C^* (I_e r_e + b r_e^2), \label{2VTObs}
\end{equation}
\noindent in which $\sigma_0$ is the central velocity dispersion of the baryonic component, $C^\ast=2\pi GC_rC_v\left(M\over L\right)_{LUM}$
and $b={2\over 3}{R\over C_r} \left(M\over L\right)_{LUM}^{-1}\rho_{DM}$, with $\left(M\over L\right)_{LUM}$ the baryonic mass-to-light ratio, and $R \equiv \langle r^2 \rangle_{LUM} / r^2_e$. For $\rho_{DM}=0$, we recover the 1VT, $\sigma_0^2=2\pi C_rC_vG M_e/r_e$, where $M_e\equiv M_{tot}/2$.

The kinematical-structural coefficients ($C_r,C_v$) translates the physical into observational quantities, and they may or may not be constant
among galaxies: $\sigma_0^2=C_v\langle v^2\rangle$ (where $\langle v^2\rangle$ is the mean square velocity of the particles), $r_e=C_rr_G$
(where $r_G$ is the gravitational radius of the system). In this work, we assume  $R/C_r\approx 20$ and $C_rC_v \approx 0.2$ as typical values for these parameters (see Paper I).

\subsection{Replotting the 2VT \label{Replot}}

In Paper I, we worked in the $\kappa$ parameter space (see \citealt{Ben92}), in which the 2VT can be expressed as

\begin{equation}
\kappa_3={\log{C^\ast}\over \sqrt{3}} + {1\over \sqrt{3}}\log{(1+b 10^\omega)}, \label{2VTKappa}
\end{equation}
\noindent where
\begin{equation}
\omega\equiv(\kappa_1 - \sqrt{3}\kappa_2)/\sqrt{2}=-\log{I_e\over r_e},
\end{equation}
\noindent and thus $\omega$ describes the central luminosity density of the stellar systems. In this space, the tilt can be understood as a systematic increase in $\kappa_3$ along the FP \citep[e.g.][]{Cio96}. Also, from Eq. \ref{2VTKappa}, we see that $\kappa_3$ depends on $C^\ast$, $b$, and $\omega$.

Now, working with directly observable quantities and recalling that the total mass-to-light ratio (refered to B-band luminosities) within the effective radius $r_e$ is given by $\log{[M_e/L_e](\odot)}=\sqrt{3}\kappa_3 - \log{C_{eff}}$ (see \citealt{Bur97}), we can rewrite the 2VT as

\begin{equation}
\log{\left[{M_e\over L_e}\right]}=\log{C^\ast} + \log{(1+ b10^\omega)} - \log{C_{eff}}
\end{equation}
\noindent or simply
\begin{equation}
\left[{M_e\over L_e}\right]=\left({C^\ast\over C_{eff}}\right)\left(1+{b\over \rho_L}\right), \label{2VTML}
\end{equation}
\noindent where $C_{eff}=2\pi G C_rC_v$ and we have defined $\rho_L\equiv I_e/r_e$ and so $\omega=\log{\rho_L^{-1}}$.

For $\rho_{DM}=0$, the only way of increasing $(M_e/L_e)$ along the FP is through a systematic variation in $C_r$ and $C_v$, which would imply that galaxies form a nonhomologous family of objects \citep{Gra97,Cap95,Cap97}. Another possibility for increasing $C^\ast$ is to suppose a systematic trend in the mass-to-light ratio with galaxy mass: $(M/L)_{LUM} \propto M^\alpha$ \citep[e.g.][]{Dre87}. However, if $\rho_{DM}\neq 0$, we can retrieve homology and still increase $M_e/L_e$ (or $\kappa_3$), depending on the variation in the ratio $b/\rho_L$, which is
the dark-to-luminous density ratio parameter of the 2VT formulation.

\begin{figure}[hptb]
\centering
\includegraphics[scale=0.4]{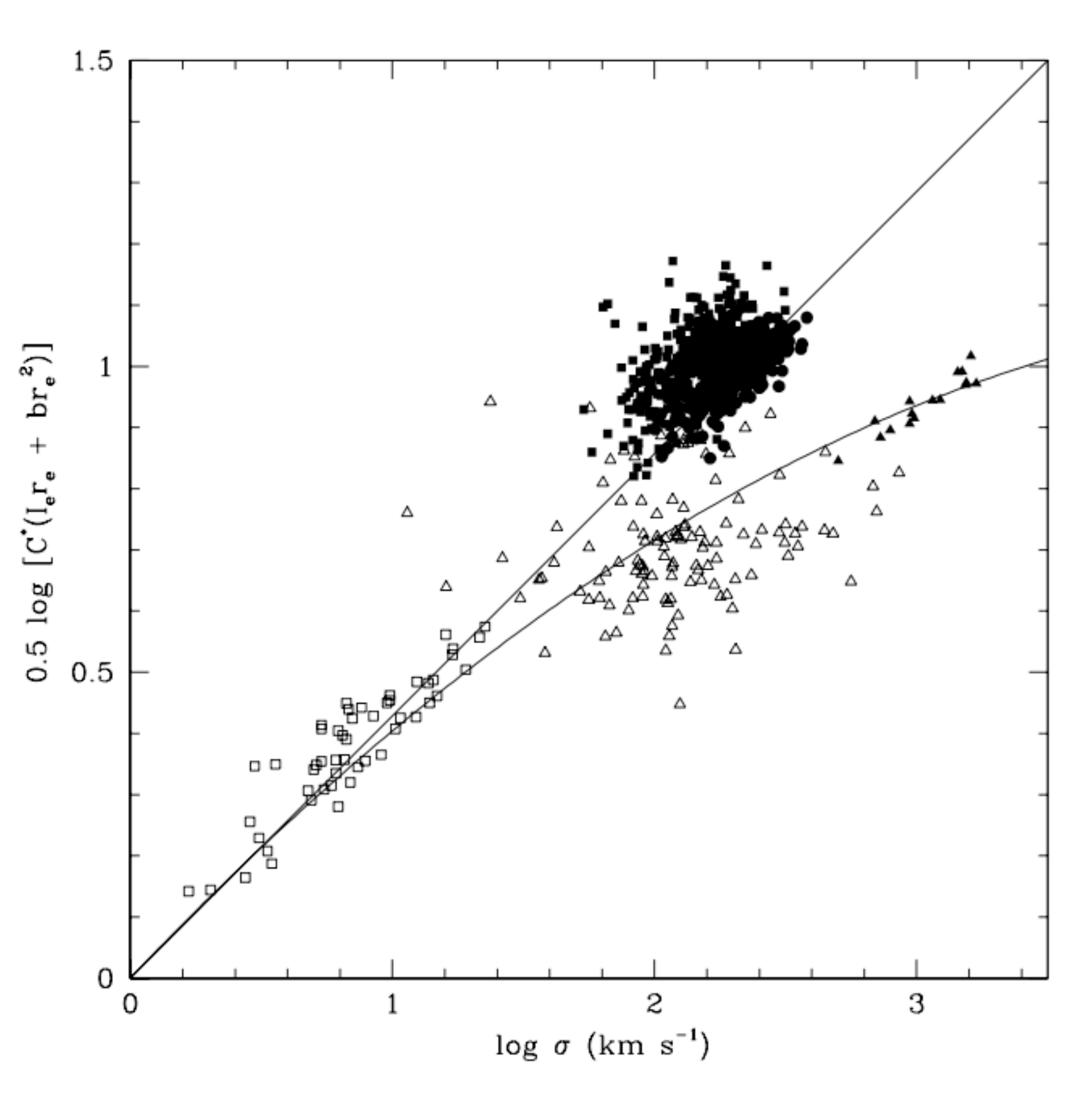}
\caption{\label{fig1} Velocity dispersion  versus
$0.5\log{[C^* (I_e r_e + b r_e^2)]}$ (in km ${\rm s^{-1}}$). The solid lines
represent the 2VT constrained by $C^{\ast}=8.28$ and $\rho_{DM}=2.3\times 10^{-2}~M_\odot {\rm pc}^{-3}$
($b=0.2$) (stellar systems) or $\rho_{DM}=5.8\times 10^{-6}~M_\odot {\rm pc}^{-3}$) ($b=4.8\times 10^{-5}$)
(galactic systems). The symbols are for globular clusters (open
squares), spiral galaxies (filled squares), 
elliptical galaxies (filled circles), groups dominated by elliptical galaxies 
(open triangles) and clusters of galaxies (filled triangles). The 1VT prediction is equivalent to the 2VT line for stellar systems.}
\end{figure}

\begin{figure}[hptb]
\centering
\includegraphics[scale=0.4]{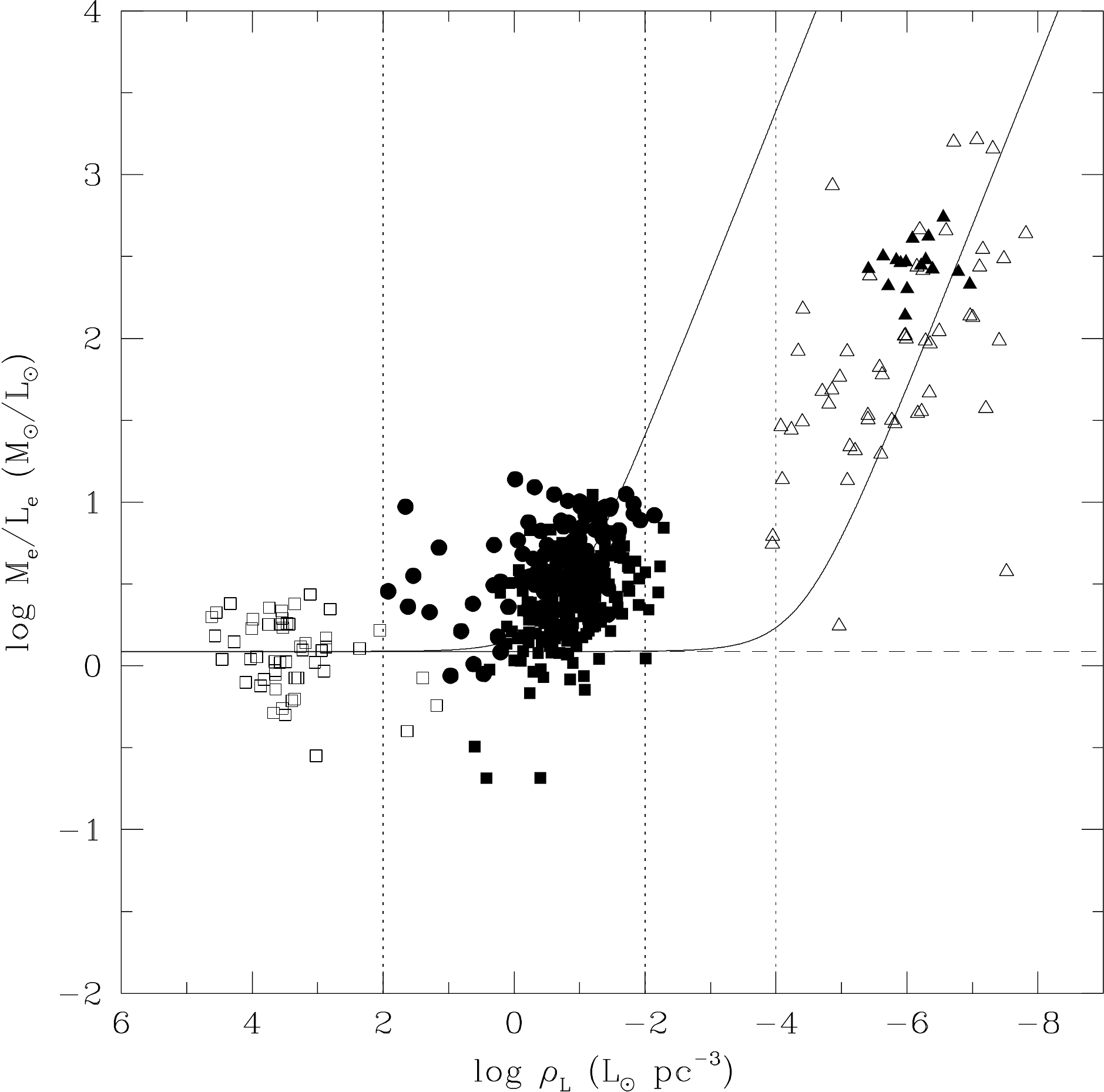}
\caption{\label{fig2} 
Effective mass-to-light ratio (in $M_\odot/L_\odot$ units) as 
a function of luminosity density ($L_\odot{\rm pc}^{-3}$). The solid lines
represent the 2VT constrained by $C^{\ast}=8.28$, $C_{eff}=6.76$, 
and $\rho_{DM}=2.3\times 10^{-2}~M_\odot {\rm pc}^{-3}$
($b=0.2$) (globular clusters and galaxies) or $\rho_{DM}=5.8\times 10^{-6}~M_\odot {\rm pc}^{-3}$ ($b=4.8\times 10^{-5}$)
(galaxy groups and clusters), while the slashed line is the 1VT.}
\end{figure}

In Paper I, a brief analysis was given for the adjusting procedure, which relies on certain assumptions about the mass-luminosity relation of the baryonic component, as well as on considerations about the DM density profile and spatial scalings for the baryonic and non-baryonic components. We refer the reader to that paper for the details. In particular, the values of $\rho_{DM}$ are found after adjusting for the parameter $b$, which gives two ``typical" values for a good global fit to the dataset, according to the type of system: globular clusters and galaxies result in $\rho_{DM} \sim 2.3\times 10^{-2}~M_\odot {\rm pc}^{-3}$, and galaxy groups and clusters favor $\rho_{DM} \sim5.8\times 10^{-6}~M_\odot {\rm pc}^{-3}$. For the present work, we assume the same adjusting assumptions (and explore the rationale behind this in a moment).  Also in accordance with Paper I, we keep $C_{eff}=6.76$ and $C^\ast=8.28$ constant (which corresponds to $(M/L)_{LUM} \sim 1.6$) for the data fittings. Based on these parameters, we plot in Figs. \ref{fig1} and \ref{fig2} the scaling relations defined by Eqs. \ref{2VTObs} and \ref{2VTML} with the observational data provided by \cite{Bur97}. We clearly see that, in the region corresponding to high central density luminosities, both 1VT and 2VT fit the points. However, in Fig. \ref{fig1}, as $\sigma$ increases, only the 2VT relation is able to fit galaxy systems. At the same time, In Fig. \ref{fig2}, as $\rho_L$ decreases, only the 2VT relation fits all the observational data. 

In the next two sections, we analyze whether the assumptions on $\rho_{DM}$ (namely, the assumed bi-modality in the dataset) are valid at the level of investigation that we aim at in the present work.  We first (Section \ref{Evidence}) quantify the evidence for the 2VT model {\it vis-\`a-vis} with the 1VT model, with the given adjustment assumptions above, using statistical arguments. Subsequently (Section \ref{Tukey}), we perform statistical pairwise comparisons in order to group systems by relative affinity in their scaling relations. 

\subsection{Evidence for 2VT \label{Evidence}}

To compare 1VT and 2VT relations, we used the Akaike information criterion \citep{Aka74}. The differences in AIC values needed for model comparison only depended on the ratio of the maximized model likelihoods and the difference in the number of parameters between the two models \citep[e.g.][]{ba02}:

\begin{equation}
AIC_1 - AIC_2 = -2\ln\left({L_1\over L_2}\right) + 2(k_1 - k_2).
\end{equation}

\noindent \cite{royall} suggests that likelihood ratios equal to 8 and 32 are equivalent to setting significance levels of 5\% and 1\%, or AIC differences of 2 and 3.5, respectively. That is, when we have $\Delta {\rm AIC}\geq 2$, the two models are significantly different.
In the present analysis, we found that data reveal a compelling difference bewteen the models, with strong evidence of 2VT. Globular clusters and galaxies have $\Delta ({\rm AIC})=2.35$ and
$\Delta ({\rm AIC})=45.90$ for the scaling relations (\ref{2VTObs}) and (\ref{2VTML}). At the same time,
galaxy groups and clusters have $\Delta ({\rm AIC})=34.31$ and
$\Delta ({\rm AIC})=18.78$ for (\ref{2VTObs}) and (\ref{2VTML}), respectively. 

\subsection{Multiple comparisons \label{Tukey}}

It is important to ascertain whether, at the present level, a refinement in the adjustments of the 2VT to the several types of systems analyzed here is an essential prerequisite for the subsequent formalism (dissipation parameter in the 2VT) that we introduce. 

Indeed, it would be desirable to have a ``minimalist" model to explain the ``cosmic metaplane" as a whole, with the least possible number of free parameters, and this was in fact the initial motivation for the 2VT formulation.  It became clear, however (see also Figs. \ref{fig1} and \ref{fig2}), that at least two ``typical" values of $\rho_{DM}$ were needed under simple assumptions. Evidently, a more thorough investigation would have to address how different assumptions on the already mentioned mass-luminosity relation of the baryonic component, the DM density profile, and spatial scalings for the baryonic and non-baryonic components affect the 2VT fittings, hence the $\rho_{DM}$ values.

The most important issue would be the strength of the main 2VT result at present, namely, that the $\rho_{DM}$ for galaxies is about $3$ orders of magnitude greater than that of groups and clusters of galaxies. It is not our intention in the present work to study the various fine-detailed adjustment values of $\rho_{DM}$ for these systems, although we acknowledge the expectation that their corresponding values could vary systematically with mass (e.g., \citealt{bull})\footnote{We aim to investigate that issue in a future work.}. What we aim at here is simply given fixed baryonic and non-baryonic density profiles, show that it is possible to formulate a further unique quantity in terms of the 2VT variables that encode dissipational information. Therefore, for the sake of our argumentation, we fix our adjustment assumptions to match those of Paper I, but acknowledge here that scaling variables, such as $\langle r^2 \rangle_{LUM}$, are actually functionals of the density profile assumed.

To certify how possible conclusions derived from our formalism  could be significantly affected by the fixed assumptions, we must analyze whether the bi-modality of $\rho_{DM}$ makes sense statistically. To answer this question, we present a pairwise comparison of means related to the variables included in relations (\ref{2VTObs}) and (\ref{2VTML}). We used the Tukey ``honest significant difference'' technique (Tukey's HSD), a quite conservative approach to getting information on the pattern of differences between the means of specific groups of objects (e.g. \citealt{hay}). We aim here to assure whether it is statistically significant to group by affinity galaxies, on one hand, and groups/clusters, on the other, from the given scaling relations of these systems, without reference to their possible specific $\rho_{DM}$ values.

\begin{figure}[hptb]
\centering
\includegraphics[scale=0.45]{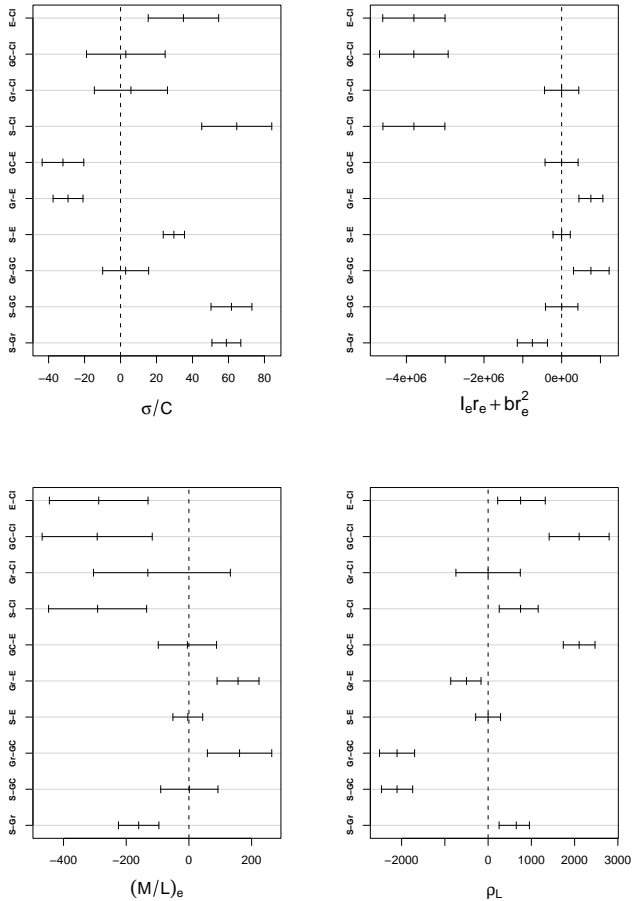}
\caption{\label{fig3}  Tukey pairwise comparisons for the scaling relations (\ref{2VTObs}) and (\ref{2VTML}) at 95\% confidence level.
Abbreviations for astrophysical systems: GC (globular clusters), S (spiral galaxies), E (elliptical galaxies), Gr (galaxy groups), and
Cl (galaxy clusters). In the upper left panel, we use $C^\ast=2\pi GC_rC_v\left(M\over L\right)_{LUM}$, instead of a constant value.}
\end{figure}

Tukey's HSD test is a method of ensuring that the  chance of finding a significant difference in any comparison is maintained at the 
confidence level of the test. For the present test, therefore, pairs of objects related by affinity in their scaling relations will have their comparison values compatible with zero. In Fig. \ref{fig3}, we see the
95\% confidence-level Tukey pairwise comparisons from our data.
In upper and lower panels we plot the differences in the variables involved in expressions (\ref{2VTObs}) and (\ref{2VTML}), respectively. The confidence intervals reveal that some differences are not significant among two or more families of objects per variable. 
These internal similarities probably set the two distinct families that lead us to the tracks in Figs. \ref{fig1} and \ref{fig2}, under the assumption of two ``typical" $\rho_{DM}$ values. In other words, pairwise comparisons allow us to
conclude that the double-adjustment found in Section \ref{Replot} is 
statistically acceptable. Therefore, for our present purposes, we regard our analysis as valid up to the point that (i) any scaling variables in the formalism are functions of the density profile, and (ii) fine-graining adjustments of the 2VT for individual systems under fixed assumptions of the density profile are not fundamental for the formalism.

\section{Dissipation formalism, 2VT, and ZOE \label{Interplay}}

While galaxies are examples of baryonic cooling and strong dissipation, groups and clusters have too long a cooling time to have dissipated a significant amount of energy on their own scale \citep[e.g.][]{Blu84}. The virial theorem (both 1VT and 2VT) cannot directly access the details of the dissipation mechanisms during structure formation. However, a slight reformulation of the 2VT allows us to build a new quantity to measure the luminous mass surface density of the self-gravitating systems, $\Sigma_{LUM}$, by relating two independent FP parameters, $\sigma_0$ and $r_e$:

\begin{equation}
\Sigma_{LUM} = {1\over C_rC_v}{\sigma_0^2\over Gr_e} - {4\pi\rho_{DM}\over 3C_r}{\langle r^2\rangle_{LUM}\over r_e}.
\end{equation}
On dimensional grounds, this quantity may be relevant in considerations of the role of dissipation in these systems.

\begin{figure}[hptb]
\centering
\includegraphics[scale=0.4]{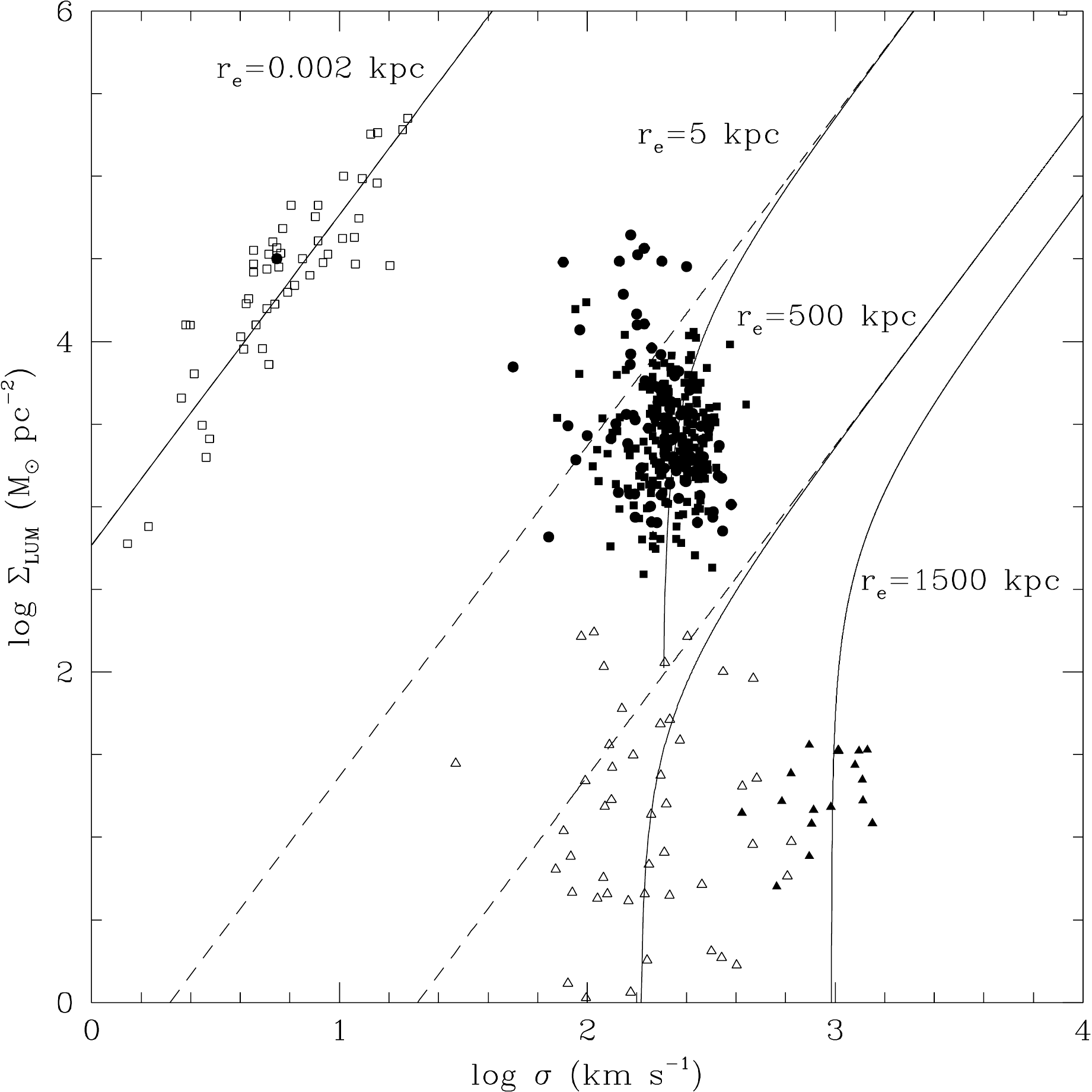}
\caption{\label{fig4} Luminous mass surface density against velocity dispersion. The solid and dashed lines are the 2VT and 1VT, respectively, for different values of $r_e$ (kpc).}
\end{figure}

In Fig. \ref{fig4}, we plot $\Sigma_{LUM}$ versus $\sigma_0$ using data from \cite{Bur97}. This projection should be visualized keeping in mind that it actually represents a projection of a higher (3D) dimensional parameter space (with $r_e$ being the third axis). Straight lines indicate the fits to the 1VT relation, whereas 2VT fits are characterized by curves that separate at some distinct points from the 1VT lines. The characteristic scales $r_e$ (assumed to be the median of the objects in each kind of system) are indicated as labels on the diagrams.

The most distinct feature of this figure is that, except for globular clusters, the 1VT implies a bad fit to data, unless one is willing to adjust an arbitrary series of 1VT relations for different $r_e$ scales in order to cover all data. On the other hand, the 2VT curves naturally bend along the $\Sigma_{LUM}$-axis and follow the data distribution, making therefore a more ``economical'' way to describe these systems. In particular, for the case of galaxy groups and clusters, typical values of $r_e$ require the 2VT bend to fit the points, while the 1VT lines would imply too large a $r_e$ for these systems.

We should also point out in this figure that there are {\it regions} defined by lines that are almost perpendicular to the 1VT lines above, and below them there are no objects at all. For instance, the so-called ``zone of exclusion" (ZOE) pointed out by \cite{Bur97} indicates an absence of high $[\sigma_0,\Sigma_{LUM}]$ objects in the Universe. Correspondingly, there is another limit to these quantities precluding the existence of very low $[\sigma_0,\Sigma_{LUM}]$ objects. The 2VT somewhat constrains this behavior since the 2VT curves tend to go asymptotically towards higher $\sigma_0$ for a given $r_e$ (in comparison to the 1VT case). A remaining question is to know how to use the virial equations to estimate the depletion of objects in certain regions of the diagram, in particular the high $[\sigma_0,\Sigma_{LUM}]$ corner. 

According to \cite{Bur97}, the equation to describe the region allowed for objects is

\begin{equation}
j_e \leq 2.95 \times 10^{14} M_e^{-4/3},
\end{equation}
\noindent where the effective luminosity density is $j_e=0.75\times 10^{-3} I_e/r_e~L_\odot {\rm pc}^{-3}$ and $\log{M_e}=\log{(\sigma_0^2 r_e)} + 5.67$. Recalling that $\omega=-\log{(I_e/r_e)}$, we have

\begin{equation}
\log{(\Sigma_e^{-1}\sigma_0^4)}\lesssim 0.75\omega + 7.52 + \log{G}.
\end{equation}

\noindent We define a new quantity to estimate the relative importance of dissipation from the equilibrium configuration of already formed astrophysical objects as

\begin{equation}
Q\equiv \left({M_e\over L_e}\right)\rho_L^{-1}=\left({C^\ast\over C_{eff}}\right)\left({1\over \rho_L} +
{b\over \rho_L^2}\right).
\end{equation}

\noindent  The idea is that, as dissipation plays less of a role, $\rho_L$ decreases while $M_e/L_e$ should increase, such that the  larger $Q$, the less dissipation that occurs; therefore, the variation in $Q$ with $\rho_L$  depicts to what degree dissipation dominates the formation process. Note that Q also depends on the  dark-to-luminous density ratio parameter $b/\rho_L$. Thus, using $Q$ and recalling that $\Sigma_e=M_e/\pi r_e^2$ and $L_e=\pi r_e^3\rho_L$, we can use use expressions (7) and (11) to find

\begin{equation}
\Sigma_e={\sigma_0^4\over C_{eff}^2\rho_L^2 Qr_e^3}.
\end{equation}

\noindent At the same time, taking $\omega=\log{\rho_L^{-1}}$ and introducing (13) into (11), we get

\begin{equation}
Q\lesssim {1\over \mathcal{C}r_e^3\rho_L^{2.75}},
\end{equation}

\noindent with $\mathcal{C}=C_{eff}^2/(3.16 \times 10^7 G)$. From this we can define

\begin{equation}
\rho_L^{max}\equiv \left({1\over Q_{min}\mathcal{C}r_e^3}\right)^{0.36}
\end{equation}

\noindent as the approximate maximum luminosity density for a given scale $r_e$. 

\begin{figure}[hptb]
\centering
\includegraphics[scale=0.4]{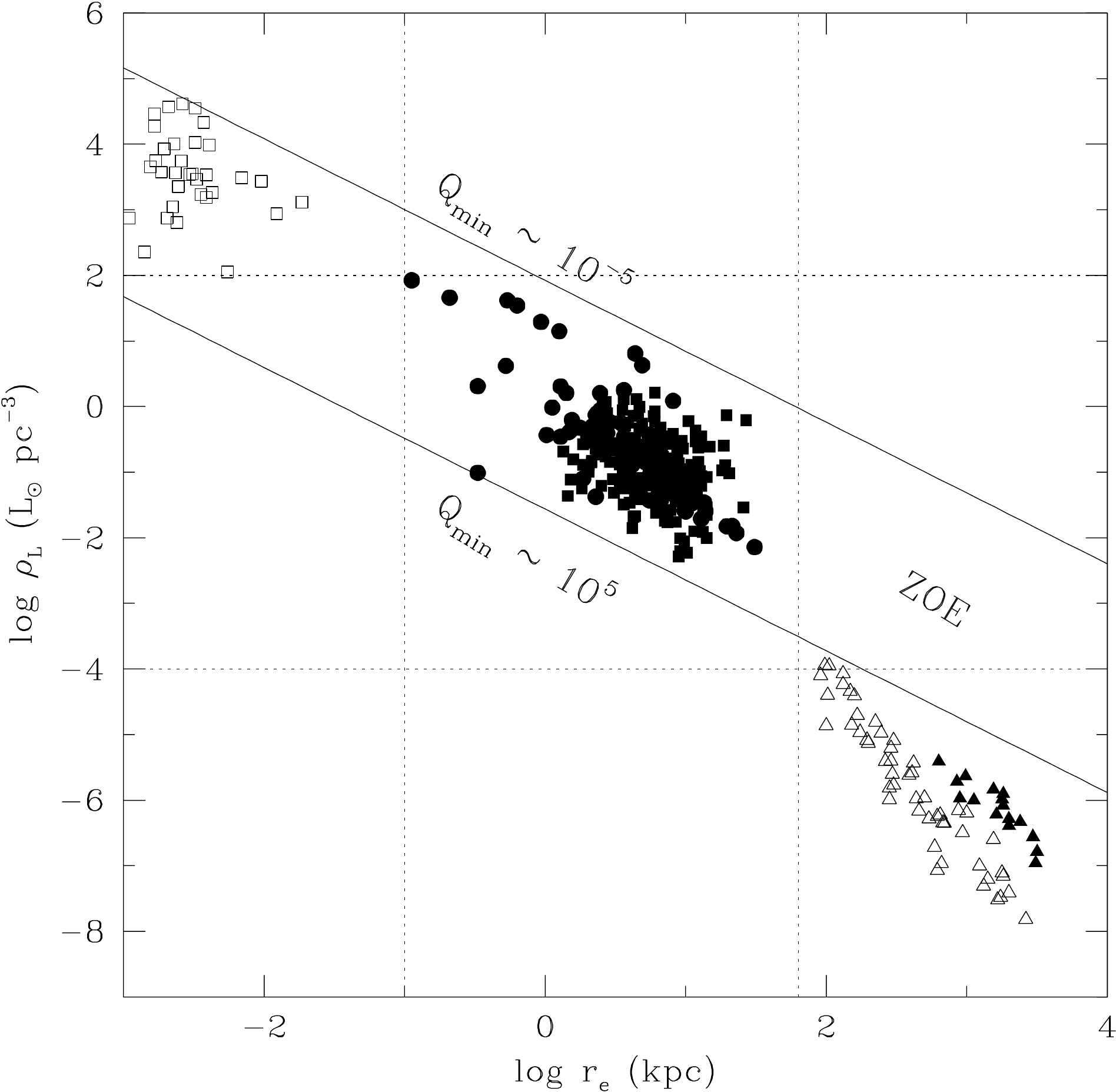}
\caption{\label{fig5} Luminosity density as a function of $r_e$. The solid lines
indicate $\rho_L^{max}$ for two values of $Q_{min}$. Dotted lines mark different locii in the plane.}
\end{figure}

The $Q_{min}$ values for the \cite{Bur97} sample are $Q_{min}^{star} \sim 10^{-5}$ (globular clusters and galaxies) and $Q_{min}^{gal}\sim 10^5~ M_\odot L_\odot^{-2}{\rm pc}^3$ (groups and clusters). Using these values in (19), we plot both $\rho_L^{max}$ and the data points on the $[r_e,\rho_L]$ plane (see Fig. \ref{fig5}). In this figure, the $Q$ lines indicate the behavior of the maximum dissipation efficiency ($\rho_L^{max}$) allowed for a given scale $r_e$. Globular clusters and galaxies have a linear relation between the two lines, while galaxy systems present a faster decrease of luminosity density
for larger radii. In fact, two values of $Q_{min}$ are necessary to regulate
the distribution of points and prevent clusters and groups from settling in the ZOE region. This result is consistent with the idea of dissipation as the main vector organizing astrophysical objects. However,  dissipation and dark matter are tightly connected (see Eq. (7)), and we necessarily have different values of central dark matter halo densities when we pass from galaxies to clusters scales. In fact, if one ignores the dark component, dissipation can be significantly overestimated at cluster scales; hence, our result suggests some influence of $\rho_{DM}$ on the dissipative infall of baryons. At the same time, the infall itself can affect the dark matter component as well. As suggested by \cite{blu86}, dissipative infall can produce smaller dark core radii and higher dark central densities. This points out the idea of a deep connection between dissipation and the final configuration of both luminous and dark distributions of matter.

\section{Discussion \label{Discussion}}

The virial metaplane with all its features encloses important aspects of galaxy formation. The phenomenology involved is rich and complex. In this work, we have concentrated our investigation on the 2VT formulation as a function of direct observables and the dissipational features on the fundamental surface defined by the 2VT as a continuation of Paper I. Our main results are

\begin{itemize}

\item Working with directly observable quantities, we fitted the 2VT
scaling relations, presented in Figs. \ref{fig1} and \ref{fig2},  for a wide range of astrophysical objects. Statistical evidence based on AIC differences and Tukey's HSD test allowed us to conclude in favor of the bi-adjust 2VT description. This result indicates the importance of dark matter to the final configuration of astrophysical objects.

\item Using the dissipational parameter $\Sigma_{LUM}$, we found two regions devoid of objects, corresponding to the upper right and lower left corners on the $[\sigma_0 ; \Sigma_{LUM} ]$ plane (Fig. \ref{fig4}). The non-existence of  low $[\sigma_0 ; \Sigma_{LUM} ]$ systems is probably related to clouds with too low temperatures and densities to form luminous objects \citep[e.g.][]{silk85}, whereas the underpopulation of high $[\sigma_0 ; \Sigma_{LUM} ]$ objects seems to agree with the work of \cite{Sil81} in the context of gas clouds collisions, which indicates a limit for the kinetic energy being lost via dissipational clustering. The results for $\Sigma_{LUM}$ (ZOE) suggest that dissipation probably plays the most important role in defining such features on the fundamental surface. 

\item Defining a new dissipational estimator, $Q$, we found an expression for the maximum luminosity density as a function of $r_e$ and $Q_{min}$ for each class of objects  (c.f. Eq. 15). Once more, the ZOE is a clear feature in the $[r_e ; \rho_L ]$ plane. Since $b$ can be written as a function of $\rho_L$ and $M_e/L_e$, this suggests a deep connection between dissipation and the final configuration of both luminous and dark distributions of matter.

\end{itemize}

Taken together, these results indicate complementary contributions of dissipation and dark matter to the orign of scaling relations in astrophysical systems. This is probably a direct consequence of the mutual influence of dark and luminous components over the dissipative collapse of these objects.

\section{Acknowledgments}
We thank B. Robertson for very useful suggestions.
We also thank H.C. Capelato and R.R. de Carvalho for interesting discussions. 
This work has the financial support of CNPq (grants 201322/2007-2 and 471254/2008-8).

\bibliographystyle{aa} 
\bibliography{Refs.bib} 

\begin{thebibliography}{47}
\expandafter\ifx\csname natexlab\endcsname\relax\def\natexlab#1{#1}\fi

\bibitem[{{Akaike}(1974)}]{Aka74}
{Akaike}, H. 1974, IEEE Trans. Autom. Contr., AC-19, 716

\bibitem[{{Araya-Melo} {et~al.}(2009){Araya-Melo}, {van de Weygaert}, \&
  {Jones}}]{Ara09}
{Araya-Melo}, P.~A., {van de Weygaert}, R., \& {Jones}, B.~J.~T. 2009, \mnras,
  400, 1317

\bibitem[{{Bekki}(1998)}]{Bek98}
{Bekki}, K. 1998, \apj, 496, 713

\bibitem[{{Bender} {et~al.}(1992){Bender}, {Burstein}, \& {Faber}}]{Ben92}
{Bender}, R., {Burstein}, D., \& {Faber}, S.~M. 1992, \apj, 399, 462

\bibitem[{{Bernardi} {et~al.}(2003){Bernardi}, {Sheth}, {Annis}, {Burles},
  {Eisenstein}, {Finkbeiner}, {Hogg}, {Lupton}, {Schlegel}, {SubbaRao},
  {Bahcall}, {Blakeslee}, {Brinkmann}, {Castander}, {Connolly}, {Csabai},
  {Doi}, {Fukugita}, {Frieman}, {Heckman}, {Hennessy}, {Ivezi{\'c}}, {Knapp},
  {Lamb}, {McKay}, {Munn}, {Nichol}, {Okamura}, {Schneider}, {Thakar}, \&
  {York}}]{Ber03}
{Bernardi}, M., {Sheth}, R.~K., {Annis}, J., {et~al.} 2003, \aj, 125, 1866

\bibitem[{{Blumenthal} {et~al.}(1986){Blumenthal}, {Faber}, {Flores}, \&
  {Primack}}]{blu86}
{Blumenthal}, G.~R., {Faber}, S.~M., {Flores}, R., \& {Primack}, J.~R. 1986,
  \apj, 301, 27

\bibitem[{{Blumenthal} {et~al.}(1984){Blumenthal}, {Faber}, {Primack}, \&
  {Rees}}]{Blu84}
{Blumenthal}, G.~R., {Faber}, S.~M., {Primack}, J.~R., \& {Rees}, M.~J. 1984,
  \nat, 311, 517

\bibitem[{{Bullock} {et~al.}(2001){Bullock}, {Kolatt}, {Sigad}, {Somerville},
  {Kravtsov}, {Klypin}, {Primack}, \& {Dekel}}]{bull}
{Bullock}, J.~S., {Kolatt}, T.~S., {Sigad}, Y., {et~al.} 2001, \mnras, 321, 559

\bibitem[{{Burnham} \& {Anderson}(2002)}]{ba02}
{Burnham}, K. \& {Anderson}, D. 2002, {Model selection and multimodel
  inference: a practical information-theoretic approach. 2nd Edition}
  (Springer-Verlag, New York, New York, USA)

\bibitem[{{Burstein} {et~al.}(1997){Burstein}, {Bender}, {Faber}, \&
  {Nolthenius}}]{Bur97}
{Burstein}, D., {Bender}, R., {Faber}, S., \& {Nolthenius}, R. 1997, \aj, 114,
  1365

\bibitem[{{Capelato} {et~al.}(1995){Capelato}, {de Carvalho}, \&
  {Carlberg}}]{Cap95}
{Capelato}, H.~V., {de Carvalho}, R.~R., \& {Carlberg}, R.~G. 1995, \apj, 451,
  525

\bibitem[{{Capelato} {et~al.}(1997){Capelato}, {de Carvalho}, \&
  {Carlberg}}]{Cap97}
{Capelato}, H.~V., {de Carvalho}, R.~R., \& {Carlberg}, R.~G. 1997, in Galaxy
  Scaling Relations: Origins, Evolution and Applications, ed. {L.~N.~da Costa
  \& A.~Renzini}, 331

\bibitem[{{Ciotti} {et~al.}(1996){Ciotti}, {Lanzoni}, \& {Renzini}}]{Cio96}
{Ciotti}, L., {Lanzoni}, B., \& {Renzini}, A. 1996, \mnras, 282, 1

\bibitem[{{Cody} {et~al.}(2009){Cody}, {Carter}, {Bridges}, {Mobasher}, \&
  {Poggianti}}]{Cod09}
{Cody}, A.~M., {Carter}, D., {Bridges}, T.~J., {Mobasher}, B., \& {Poggianti},
  B.~M. 2009, \mnras, 396, 1647

\bibitem[{{Coles} \& {Lucchin}(1995)}]{Col95}
{Coles}, P. \& {Lucchin}, F. 1995, {Cosmology. The origin and evolution of
  cosmic structure} (Chichester: Wiley, |c1995)

\bibitem[{{Dantas} {et~al.}(2000){Dantas}, {Ribeiro}, {Capelato}, \& {de
  Carvalho}}]{Dan00}
{Dantas}, C.~C., {Ribeiro}, A.~L.~B., {Capelato}, H.~V., \& {de Carvalho},
  R.~R. 2000, \apjl, 528, L5

\bibitem[{{De Lucia} \& {Blaizot}(2007)}]{DeL07}
{De Lucia}, G. \& {Blaizot}, J. 2007, \mnras, 375, 2

\bibitem[{{Djorgovski} \& {Davis}(1987)}]{Djo87}
{Djorgovski}, S. \& {Davis}, M. 1987, \apj, 313, 59

\bibitem[{{Djorgovski} \& {de Carvalho}(1990)}]{Djo90}
{Djorgovski}, S. \& {de Carvalho}, R. 1990, in Astrophysics and Space Science
  Library, Vol. 160, Windows on Galaxies, ed. {G.~Fabbiano, J.~S.~Gallagher, \&
  A.~Renzini}, 9

\bibitem[{{Dodelson}(2003)}]{Dod03}
{Dodelson}, S. 2003, {Modern cosmology} (Amsterdam (Netherlands): Academic
  Press)

\bibitem[{{Dressler} {et~al.}(1987){Dressler}, {Lynden-Bell}, {Burstein},
  {Davies}, {Faber}, {Terlevich}, \& {Wegner}}]{Dre87}
{Dressler}, A., {Lynden-Bell}, D., {Burstein}, D., {et~al.} 1987, \apj, 313, 42

\bibitem[{{Fritsch} \& {Buchert}(1999)}]{Fri99}
{Fritsch}, C. \& {Buchert}, T. 1999, \aap, 344, 749

\bibitem[{{Gadotti}(2009)}]{Gad09}
{Gadotti}, D.~A. 2009, \mnras, 393, 1531

\bibitem[{{Gao} \& {Theuns}(2007)}]{Gao07}
{Gao}, L. \& {Theuns}, T. 2007, Science, 317, 1527

\bibitem[{{Graham} \& {Colless}(1997)}]{Gra97}
{Graham}, A. \& {Colless}, M. 1997, \mnras, 287, 221

\bibitem[{{Guzman} {et~al.}(1993){Guzman}, {Lucey}, \& {Bower}}]{Guz93}
{Guzman}, R., {Lucey}, J.~R., \& {Bower}, R.~G. 1993, \mnras, 265, 731

\bibitem[{{Hayter}(1986)}]{hay}
{Hayter}, A.~J. 1986, Journal of the American Statistical Association, 81, 1001

\bibitem[{{Hjorth} \& {Madsen}(1995)}]{Hjo95}
{Hjorth}, J. \& {Madsen}, J. 1995, \apj, 445, 55

\bibitem[{{Hopkins} {et~al.}(2008){Hopkins}, {Cox}, \& {Hernquist}}]{Hop08}
{Hopkins}, P.~F., {Cox}, T.~J., \& {Hernquist}, L. 2008, \apj, 689, 17

\bibitem[{{Jablonka} {et~al.}(1996){Jablonka}, {Martin}, \& {Arimoto}}]{Jab96}
{Jablonka}, P., {Martin}, P., \& {Arimoto}, N. 1996, \aj, 112, 1415

\bibitem[{{Kritsuk}(1997)}]{Kri97}
{Kritsuk}, A.~G. 1997, \mnras, 284, 327

\bibitem[{{Lanzoni} {et~al.}(2004){Lanzoni}, {Ciotti}, {Cappi}, {Tormen}, \&
  {Zamorani}}]{Lan04}
{Lanzoni}, B., {Ciotti}, L., {Cappi}, A., {Tormen}, G., \& {Zamorani}, G. 2004,
  \apj, 600, 640

\bibitem[{{Lanzoni} {et~al.}(2005){Lanzoni}, {Ciotti}, {Cappi}, {Tormen}, \&
  {Zamorani}}]{Lan05}
{Lanzoni}, B., {Ciotti}, L., {Cappi}, A., {Tormen}, G., \& {Zamorani}, G. 2005,
  in Multiwavelength Mapping of Galaxy Formation and Evolution, ed. {A.~Renzini
  \& R.~Bender}, 414

\bibitem[{{McIntosh} {et~al.}(2008){McIntosh}, {Guo}, {Hertzberg}, {Katz},
  {Mo}, {van den Bosch}, \& {Yang}}]{McI08}
{McIntosh}, D.~H., {Guo}, Y., {Hertzberg}, J., {et~al.} 2008, \mnras, 388, 1537

\bibitem[{{Padmanabhan}(2006)}]{Pad06}
{Padmanabhan}, T. 2006, in American Institute of Physics Conference Series,
  Vol. 843, Graduate School in Astronomy: X, ed. {S.~Daflon, J.~Alcaniz,
  E.~Telles, \& R.~de la Reza}, 111--166

\bibitem[{{Pahre} {et~al.}(1995){Pahre}, {Djorgovski}, \& {de
  Carvalho}}]{Pah95}
{Pahre}, M.~A., {Djorgovski}, S.~G., \& {de Carvalho}, R.~R. 1995, \apjl, 453,
  L17

\bibitem[{{Pahre} {et~al.}(1998){Pahre}, {Djorgovski}, \& {de
  Carvalho}}]{Pah98}
{Pahre}, M.~A., {Djorgovski}, S.~G., \& {de Carvalho}, R.~R. 1998, \aj, 116,
  1591

\bibitem[{{Peebles}(1980)}]{Pee80}
{Peebles}, P.~J.~E. 1980, {The large-scale structure of the universe}
  (Princeton, N.J., Princeton University Press)

\bibitem[{{Proctor} {et~al.}(2008){Proctor}, {Lah}, {Forbes}, {Colless}, \&
  {Couch}}]{Pro08}
{Proctor}, R.~N., {Lah}, P., {Forbes}, D.~A., {Colless}, M., \& {Couch}, W.
  2008, \mnras, 386, 1781

\bibitem[{{Robertson} {et~al.}(2006){Robertson}, {Cox}, {Hernquist}, {Franx},
  {Hopkins}, {Martini}, \& {Springel}}]{Rob06}
{Robertson}, B., {Cox}, T.~J., {Hernquist}, L., {et~al.} 2006, \apj, 641, 21

\bibitem[{{Royall}(1997)}]{royall}
{Royall}, R. 1997, {Statistical Evidence a Likelihood Paradigm} (Chapman and
  Hall/CRC)

\bibitem[{{Schaeffer} {et~al.}(1993){Schaeffer}, {Maurogordato}, {Cappi}, \&
  {Bernardeau}}]{Sch93}
{Schaeffer}, R., {Maurogordato}, S., {Cappi}, A., \& {Bernardeau}, F. 1993,
  \mnras, 263, L21

\bibitem[{{Schaeffer} \& {Silk}(1985)}]{silk85}
{Schaeffer}, R. \& {Silk}, J. 1985, \apj, 292, 319

\bibitem[{{Silk} \& {Norman}(1981)}]{Sil81}
{Silk}, J. \& {Norman}, C. 1981, Ap. J., 247, 59

\bibitem[{{Tortora} {et~al.}(2009){Tortora}, {Napolitano}, {Romanowsky},
  {Capaccioli}, \& {Covone}}]{Tor09}
{Tortora}, C., {Napolitano}, N.~R., {Romanowsky}, A.~J., {Capaccioli}, M., \&
  {Covone}, G. 2009, \mnras, 396, 1132

\bibitem[{{Trujillo} {et~al.}(2004){Trujillo}, {Burkert}, \& {Bell}}]{Tru04}
{Trujillo}, I., {Burkert}, A., \& {Bell}, E.~F. 2004, \apjl, 600, L39

\bibitem[{{Zaritsky} {et~al.}(2006){Zaritsky}, {Gonzalez}, \&
  {Zabludoff}}]{Zar06}
{Zaritsky}, D., {Gonzalez}, A.~H., \& {Zabludoff}, A.~I. 2006, \apj, 638, 725

\end{thebibliography}

\end{document}